\documentclass[prb,twocolumn,amsmath,amssymb,showpacs]{revtex4}

\usepackage{graphicx}
\usepackage{dcolumn}
\usepackage{bm}

\newcommand{\lco}{LaCoO$_3$}
\newcommand{\lsco}{La$_{1-x}$Sr$_x$CoO$_3$}
\newcommand{\lscod}{La$_{1-x}$Sr$_x$CoO$_{3+\delta}$}

\newcommand{\nco}{Na$_x$CoO$_2$}

\newcommand{\ch}{susceptibility}

\newcommand{\ka}{$\kappa$}
\newcommand{\kch}{$\kappa_{ch}$}
\newcommand{\kph}{$\kappa_{ph}$}

\newcommand{\tg}{$t_{2g}$}
\newcommand{\eg}{$e_{g}$}

\begin{document}

\title{Thermal Conductivity, Thermopower, and Figure of Merit of \lsco}

\author{K.~Berggold}
\author{M.~Kriener}
\author{C.~Zobel}
\author{A.~Reichl}
\author{M.~Reuther}
\author{R.~M\"{u}ller}
\author{A.~Freimuth}
\author{T.~Lorenz}
\affiliation{ II.~Physikalisches Institut, Universit\"{a}t
zu K\"{o}ln, Z\"{u}lpicher Str. 77, 50937 K\"{o}ln, Germany}

\date{\today}

\begin{abstract}

We present a study of the thermal conductivity $\kappa$ and the
thermopower $S$ of single crystals of \lsco\ with $0\le x \le
0.3$. For all Sr concentrations \lsco\ has rather low $\kappa$
values. For the insulators ($x<0.18$) this arises from a
suppression of the phonon thermal conductivity by lattice disorder
due to temperature- and/or doping-induced spin-state transitions
of the Co ions. For larger $x$, the heat transport by phonons
remains low, but an additional contribution from mobile charge
carriers causes a moderate increase of $\kappa $. The thermopower
of the low-doped crystals is positive and shows a pronounced
maximum as a function of temperature. With increasing $x$, this
maximum strongly broadens and its magnitude decreases. For the
highest Sr content ($x=0.3$) $S$ becomes even negative in the
intermediate temperature range. From $S$, $\kappa$, and the
electrical resistivity $\rho$ we derive the thermoelectric figure
of merit $Z=S^2\,/\,\kappa\rho$.  For intermediate Sr
concentrations we find notably large values of $Z$ indicating
that Co-based materials could be promising candidates for
thermoelectric cooling.

\end{abstract}

\pacs{72.15.Jf, 72.80.Ga, 71.30.+h}


\maketitle

\section{Introduction}

Among transition-metal oxides cobalt compounds are of particular
interest due to the possibility of the Co ions to occur in
different spin states. This is a long standing issue for the
almost cubic perovskite \lco\ where the Co$^{3+}$ spin state
changes as a function of temperature from a non-magnetic low-spin
(LS, S\,=\,0) state to an intermediate-spin (IS, S\,=\,1) or a
high-spin (HS, S\,=\,2) state, which both are
magnetic.\cite{jonker53a,asai94a,senaris95a,korotin96a,yamaguchi97a,asai98a,zobel02a}
During the last years cobaltates with layered CoO structures have
also become subject of intense
studies.\cite{vogt00a,suard00a,moritomo00a,respaud01a,wu01a,wang01a,roy02a,hu04a}
It has been proposed that various of these compounds would also
show temperature-dependent spin-state transitions of the
Co$^{3+}$ and/or Co$^{2+}$ ions, but an unambiguous proof of such
spin-state transitions is still missing (see e.\,g.\
Ref.\,\onlinecite{hu04a}). Recently, the observation of
superconductivity in Na$_x$CoO$_2\cdot y$\,H$_2$O has attracted
much attention.\cite{takada03a} The water-free parent compound
\nco\ became prominent some years ago already in a different
context.\cite{takahata00a} It was found that \nco\ with $x=0.6$
has a metallic electrical conductivity $\sigma$, but a low
thermal conductivity $\kappa$ and, in addition, a large
thermopower $S$. The combination of large $\sigma$, small
$\kappa$, and large $S$ values is a precondition for effective
thermoelectric cooling. The performance of thermoelectric devices
depends on the so-called thermoelectric figure of merit
$Z=S^2\,/\,\kappa\rho$ where $\rho=1/\sigma$ denotes the
electrical resistivity. For an effective cooling, $ZT$ values
($T$ is the absolute temperature) of order unity should be
reached and are found for instance in Bi-based alloys, some
thin-film devices or quantum dot
superlattices.\cite{venkata_nature01a,harmann_science02a} For
comparison, typical metals have much smaller $ZT$ values of order
$10^{-4}$. In this respect it was quite surprising that \nco\ has
$ZT \simeq 0.03$ for $150\,{\rm K}\le T\le 300$\,K. The enhanced
figure of merit of \nco\ mainly arises from an enhanced
thermopower.\cite{terasaki97a} Based on a study of the
magnetic-field dependence $S(H)$ of \nco\ it has been argued that
the spin entropy is the likely source for the large
thermopower.\cite{wang03a}

A large thermopower occurs also in \lsco .\cite{senaris95b} This
motivated us to study the transport properties of this series in
order to determine experimentally the $ZT$ values and their
dependence on temperature and doping. Another motivation for our
study was that the previous results do not give a consistent
picture of the transport properties. For example, from various
anomalous features in the temperature dependencies of $S$,
$\rho$, and the magnetic \ch\ $\chi$ a complex phase diagram was
proposed for the \lsco\ system.\cite{senaris95b} It has, however,
been argued that the occurrence of some of these anomalous
features depends on the preparation technique of the
polycrystals.\cite{anilkumar98a} In fact, the phase diagram
derived from $\rho(T)$ and $\chi(T)$ measured on \lsco\ single
crystals\,\cite{itoh94a,kriener04a} is much less complex than the
previous one\,\cite{senaris95b}. Moreover, the temperature
dependencies of $S$ for undoped \lco\ reported in
Refs.\,\onlinecite{senaris95a} and~\onlinecite{sehlin95a} are
contradictory. For $T>400$\,K both reports find a positive
thermopower of order $+50\,\mu$V/K, which slightly increases with
decreasing temperature. For $T<400$\,K, however, a further
increase of $S$ with a maximum of about $+1200\,\mu$V/K around
100\,K is found in Ref.\,\onlinecite{senaris95a}, whereas a sign
change of $S$ and a decrease to about $-400\,\mu$V/K for $T\simeq
200$\,K is reported in Ref.\,\onlinecite{sehlin95a}. The thermal
conductivity of the \lsco\ system has to our knowledge not yet
been studied at all.

In this report we present an experimental study of the thermal
conductivity and the thermopower of a series of single crystals
of \lsco\ with $0\le x\le 0.3$. The paper is organized as follows:
In the next section we briefly describe the experimental setup and
introduce the crystals, which (partly) have already been used in
previous studies.\cite{zobel02a,kriener04a,baier05a,lengsdorf04a}
In section\,\ref{resdis} our measurements of $\kappa$ and $S$ are
discussed and we derive the thermoelectric figure of merit for
the \lsco\ series. The main results are summarized in the last
section.

\section{Experimental}

The single crystals used in this study have been grown by a
floating-zone technique in an image furnace. Details of the sample
preparation and their characterization by X-ray diffraction,
magnetization, and resistivity measurements are given in
Ref.\,\onlinecite{kriener04a}. At low temperatures \lco\ is a good
insulator, but with increasing Sr content the resistivity
systematically decreases and \lsco\ becomes metallic for $x>0.18$.
Samples with Sr concentrations above the metal insulator
transition show ferromagnetic order with transition temperatures
$T_c$ around 220\,K, which slightly increase with $x$. The
magnetic order is accompanied by a kink-like decrease of
$\rho(T)$ below $T_c$.\cite{kriener04a}

The nominal valence of the Co ions does not only depend on the Sr
content, but also on the oxygen concentration, i.\,e.\ the amount
$n$ of charge carriers in \lscod\ is given by $n=x+2\delta$.
Positive (negative) values of $n$ mean hole (electron) doping and
formally an amount $n$ of the Co$^{3+}$ ions is transformed into
Co$^{4+}$ (Co$^{2+}$).\cite{remarkO2p,blockade} Because of the
high oxidation state of Co$^{4+}$, one may suspect that with
increasing Sr concentration the amount of oxygen vacancies also
increases, i.\,e.\ an increasing $x$ could be partially
compensated by a decreasing $\delta $. In order to check this we
have determined the oxygen content of the entire series \lscod\ by
thermogravimetric analysis (TGA/SDTA851, Mettler-Toledo). Pieces
of about 50\,mg of the single crystals have been ground and
heated up to 900$^\circ$C in a reducing atmosphere (N$_2$ with
5\,\% H$_2$) in order to decompose \lscod\ into La$_2$O$_3$, SrO
and elementary Co.\cite{james04a,goossens04a,conder05a} The value
of $\delta $ is then calculated from the measured weight loss. We
have tested the reproducibility of our method by repeatedly
measuring amounts of about 50\,mg from the same badge of a \lco\
polycrystal. The different results agree to each other within
$\pm 0.01$, which is comparable to the uncertainty of the same
and alternative methods of oxygen-content determination in
cobaltates.\cite{james04a,goossens04a,conder05a} A scatter of $\pm
0.01$ is also present in the determined oxygen contents of our
\lscod\ crystals. A linear fit of $\delta(x)$ yields a weak
decrease ${\rm d\delta /d}x=-0.05$. Therefore we determine the
charge carrier content via $n(x)=x-0.1\cdot x$, i.\,e.\ our
analysis reveals a 10\% reduction of the charge-carrier content
with respect to the Sr content. We do not calculate $n=x+2\delta$
for each crystal individually, because the scatter of $\delta$
would correspond to a scatter of the charge carrier content of
$\pm 0.02$, which can be excluded for the studied crystals from
the measurements of the magnetization and resistivity (see
Ref.\,\onlinecite{kriener04a}). Both quantities vary monotonously
as a function of $x$ for $0\le x \le 0.3$. Thus the scatter of $n$
between samples with neighboring $x$ is much smaller than their
difference $\Delta x$, which amounts e.\,g.\ to only 0.01 for the
lowest concentrations. This conclusion is also confirmed by the
data presented in this report. The only exception is the
thermopower of nominally undoped \lco , which will be discussed
below.

The measurements of thermal conductivity and thermopower have
been performed by a steady-state technique. One end of the
rectangular bar-shaped crystals of typical dimensions $1\times
2\times 4$\,mm$^3$ has been glued on a temperature-stabilized Cu
plate using an insulating varnish (VGE-7031, LakeShore). A small
resistor has been attached to the other end of the sample in
order to produce a heat current through the sample. The
corresponding temperature gradient over the sample was measured
by a differential Chromel-Au+0.07\%Fe or a Chromel-Constantan
thermocouple which has also been glued to the sample. Typical
temperature gradients were of the order of 0.2\,K. In addition,
two gold contacts have been evaporated on the sample surface and
two copper leads have been connected to these contacts by
conductive silver epoxy in order to determine the thermopower.
All offset voltages and the thermopower of the Copper leads have
been carefully subtracted. With our setup we do not obtain
reliable results for $S$ on insulating samples with resistivities
of the order of 10$^{9}\,\Omega$cm or larger, as it is the case
for \lsco\ with $x\le 0.002$ below about 100\,K. The absolute
accuracy of the thermal conductivity and thermopower measurements
is restricted to about 10\% by uncertainties in the sample
geometry, whereas the relative accuracy is of the order of a few
percent.

\section{Results and Discussion}
\label{resdis}

\subsection{Thermal Conductivity}

\begin{figure}
\includegraphics[angle=0,width=8cm]{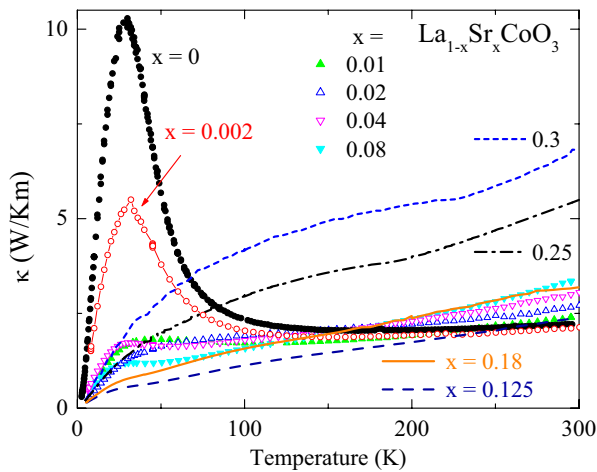}
 \caption[]{Thermal conductivity of \lsco\ as a  function of temperature
 for different doping $x$.} \label{wlf}
\end{figure}

In Fig.\,\ref{wlf} we show the thermal conductivity of \lsco . We
concentrate first on the undoped compound. In an insulator like
\lco\ the heat is usually carried by acoustic phonons. Optical
phonons play a minor role since their dispersion is typically
much weaker than that of the acoustic branches. A qualitative
understanding of the phononic heat transport is obtained from the
relation $\kappa \propto cv\ell$ where $c$, $v$, and $\ell$ denote
the specific heat, the sound velocity and the mean free path of
the phonons, respectively. At low temperatures $\ell$ is
determined by lattice imperfections, or by the sample dimensions
in very clean crystals, and the temperature dependence of $\kappa$
follows that of the specific heat, i.e. $\kappa \propto c \propto
T^3$ for $T\rightarrow 0$\,K. At intermediate temperatures a
maximum of \ka\ occurs, when phonon-phonon Umklapp scattering
becomes important and leads to a decrease of $\ell$ with
increasing temperature. At high temperatures $\ell$ is
proportional to the number of phonons, i.e. $\ell \propto 1/T$
and $\kappa$ follows roughly a $1/T$ dependence if the temperature
dependence of $c$ is not too strong, which is the case when the
Debye temperature is approached.

At first glance the results on \lco\ seem to be consistent with
usual phononic heat transport. However, a closer inspection of the
data reveals several anomalous features. The thermal conductivity
above about 100\,K  is anomalously small and its temperature
dependence is very unusual, since above 150\,K the thermal
conductivity slightly increases with increasing temperature. The
small absolute values cannot be attributed to strong defect
scattering, since conventional defect scattering does not vanish
below 100\,K so that \ka\ would not increase and show a maximum
below 100\,K. Therefore, the small \ka\ and its unusual
temperature dependence indicate that an unusual, additional
scattering mechanism is active above about 25\,K. It is, in
principle, possible that low-lying optical phonons cause resonant
scattering of acoustic phonons and therefore suppress the heat
current in a certain temperature range.\cite{verweis_scbo} This
could be an intrinsic feature of the phonons of the cubic
perovskite structure or soft phonon modes could arise from a
structural instability. However, there is, to our knowledge, no
evidence for one of these scenarios. Thus we regard an anomalous
damping by optical phonons as rather unlikely.

In \lco\ there is another source for an unusual heat conductivity,
which is related to the temperature-induced spin-state transition.
With increasing temperature a thermal population of the IS (or HS)
state takes place. As has been shown in
Refs.\,\onlinecite{zobel02a,baier05a} this causes a strongly
anomalous thermal expansion above about 25\,K. A microscopic
picture of the anomalous thermal expansion is obtained from the
fact that in the LS state all six 3d~electrons are in the \tg\
orbitals, whereas in the IS or HS states also \eg\ orbitals are
occupied. Since the \eg\ orbitals point towards the O$^{2-}$ ions
their population will expand the corresponding CoO$_6$ octahedra.
Above 25\,K \lco\ will therefore contain smaller and larger
CoO$_6$ octahedra with Co$^{3+}$ ions in the LS and IS (HS)
state, respectively. This implies an increasing lattice disorder
above 25\,K, and therefore an additional suppression of the
thermal conductivity. The disorder and thus the suppression of
\ka\ is most pronounced for an equal number of LS and IS (HS)
ions. For the proposed LS to IS state scenario of \lco\ with an
energy gap $\Delta\simeq 180$\,K between the LS and IS states an
equal occupation of LS and IS ions is expected around $T\simeq
165$\,K. Therefore the strong reduction of \ka\ above 25\,K, its
small absolute values and the weak minimum around 150\,K can
arise from temperature-induced LS-IS disorder. This picture,
which has also been proposed in Ref.\,\onlinecite{yan04a}, yields
a consistent interpretation of the thermal conductivity data of
\lco . We note that the spin-state transition appears very
favorable for the search of large $ZT$ values, because it may
intrinsically suppress \ka . A more quantitative study of the
influence of the spin-state transition on \ka\ is currently
underway.

The thermal conductivity of the crystal with $x=0.002$ is similar
to that of pure \lco , but the low-temperature maximum of
$\kappa$ is already strongly suppressed. For higher Sr doping
this maximum is almost completely absent and $\kappa$ increases
continuously with increasing temperature. The room temperature
values of $\kappa$ lie between 2 and 3\,W/Km for all crystals
with $x\le 0.18$. We attribute the drastic suppression of $\kappa$
at low temperatures to a Sr-induced disorder, which hinders a
strong increase of $\ell$ for $T\rightarrow 0$\,K. Probably this
disorder does not solely arise from the bare difference between
La$^{3+}$ and Sr$^{2+}$ ions. From magnetization measurements it
is found that for $x\le 0.01$ so-called magnetic polarons with
high spin values ($S=10-16$) are formed.\cite{yamaguchi96a} The
idea is, that the divalent Sr$^{2+}$ ions nominally cause the same
amount of Co$^{4+}$ ions, and that these magnetic ions induce a
spin-state transition of the neighboring Co$^{3+}$ ions from LS
to IS or HS states. Due to such a polaron formation the disorder
is strongly enhanced for the lowest Sr concentrations, whereas
for larger $x$ the polarons start to overlap and the enhancement
becomes less effective.

Samples with $x>0.18$ show metallic conductivity and
ferromagnetic order at low temperatures. In this concentration
range one expects that magnetic polarons become much less
important. However, the low-temperature peak of \ka\ remains
absent, since the Sr concentration is so large, that the bare
doping-induced lattice disorder is sufficient to suppress the
low-temperature peak. Moreover, the Sr doping induces mobile
charge carriers, which serve as additional scatterers for the
phonons.

The mobile charge carriers for larger $x$ are expected to
transport heat, too. The total thermal conductivity therefore
consists of a phononic contribution and a contribution of mobile
charge carriers, i.\,e.\ $\kappa = \kappa_{ph} +\kappa_{ch}$.
Usually, $\kappa_{ch}$ can be estimated by the Wiedemann-Franz
law, which relates $\kappa_{ch}$ to the electrical conductivity
according to $\kappa_{ch} \simeq L_0\,\sigma\, T$. Here,  $L_0=
2.44\cdot 10^{-8}$\,V$^2$/K$^2$ denotes the Sommerfeld value of
the Lorenz number. From the room temperature values of $\rho$ we
estimate $\kappa_{ch}(300$\,K$)\simeq 0.4$, 0.9, 1.6 and 2.7\,W/Km
for $x=0.125$, 0.18, 0.25 and 0.3, respectively, and
$\kappa_{ch}\ll \kappa$ for smaller $x$. Additional heat
conduction by charge carriers is therefore relevant for metallic
\lsco\ only and explains why the crystals with $x\ge 0.25$ have
significantly larger $\kappa(T)$ values for $T> 100$\,K than the
insulating samples with $x<0.18$.

\begin{figure}
\includegraphics[angle=0,width=8cm]{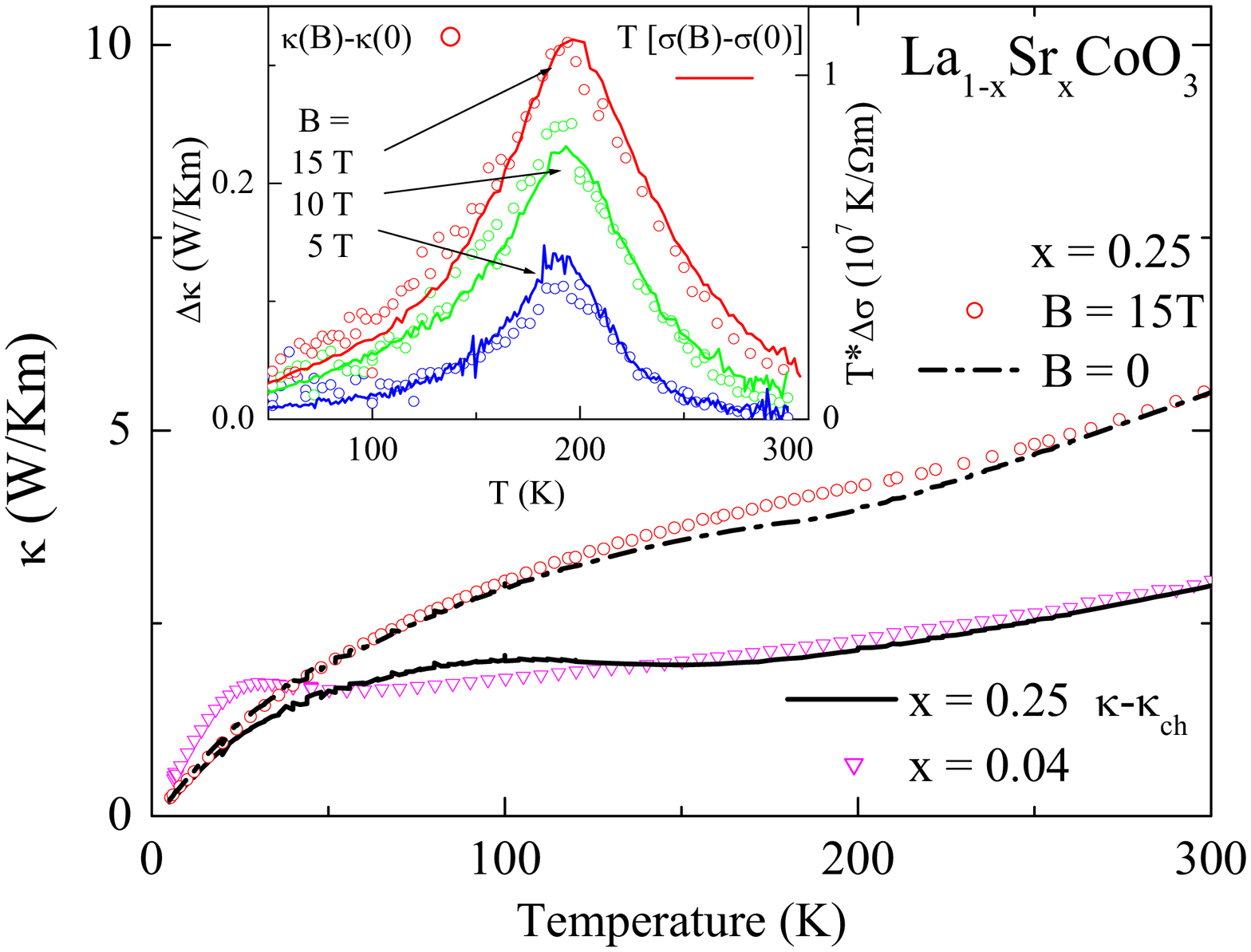}
 \caption[]{Thermal conductivity \ka\ of \lsco\ with $x=0.25$ as a
 function of temperature measured in 15\,T ($\circ$) and in zero
 magnetic field ($-\cdot -$). The solid line is the phononic
 contribution \kph\ of the $x=0.25$ crystal, which is obtained by
 $\kappa_{ph}= \kappa(B) - L\,T\,\sigma(B)$ using the
 electrical conductivity $\sigma$ and the Lorenz number $L$.
 For comparison $\kappa$ of the $x=0.04$ sample ($\triangledown$)
 is also depicted. The Inset shows the magnetic-field dependencies
 $\Delta\kappa(B)=\kappa(B)-\kappa(B=0)$ ($\circ$, left $y$ scale)
 and $T\,\Delta\sigma(B)=T\,\left[\sigma(B)-\sigma(B=0)\right]$
 (---, right $y$ scale) for $B=5$, 10, and 15\,T, respectively.
 The value of the Lorenz number is obtained from the field- and
 temperature-independent scaling factor between both quantities,
 i.\,e.\ via $L= \Delta\kappa(B)\,/\,\left[T\,\Delta\sigma(B)\right]
 =2.9\cdot 10^{-8}$\,V$^2$/K$^2$.} \label{kvb}
\end{figure}

For $x=0.25$ and $0.3$ anomalies of \ka\ occur around 200\,K and
230\,K, respectively, i.\,e.\ close to the respective
ferromagnetic ordering temperatures $T_c$. Below $T_c$, \ka\
increases and we attribute this to the decrease of $\rho$ below
$T_c$,\cite{kriener04a} which leads to a corresponding increase
of \kch . Since the charge transport close to $T_c$ depends on a
magnetic field in \lsco ,\cite{mahendiran96a} we may analyze
\kch\ for $x=0.25$ in more detail by comparing the magnetic-field
dependencies of $\kappa$ and $\sigma=1/\rho$ (Fig.\,\ref{kvb}). We
find an increase of \ka\ with increasing field, that is most
pronounced around 200\,K where the strongest magnetic-field
induced suppression of $\rho$ is observed,
too.\cite{mahendiran96a} Under the reasonable assumption of a
negligible field dependence of \kph\ we obtain $\Delta\kappa
(T,B)= \kappa (T,B)-\kappa (T,0)=\kappa_{ch}(T,B)-\kappa_{ch}
(T,0)$, and the Lorenz number is given by the relation
$L=\Delta\kappa (T,B)\,/\,T\Delta\sigma(T,B)$ with
$\Delta\sigma(T,B)= \sigma (T,B)-\sigma(T,0)$. As shown in the
inset of Fig.\,\ref{kvb} the scaling relation between
$\Delta\kappa$ and $T\Delta\sigma(T,B)$ is well fulfilled over
the entire temperature and magnetic-field range studied here.
From this scaling we find $L=2.9\cdot 10^{-8}$\,V$^2$/K$^2$,
which is about 20\% larger than $L_0$. The phononic thermal
conductivity for $x=0.25$ is then obtained by $\kappa_{ph}=
\kappa(B) - L\,T\,\sigma(B)$ and found to agree well with those
of the low-doped samples.

\subsection{Thermopower}

\begin{figure}
\includegraphics[angle=0,width=8cm]{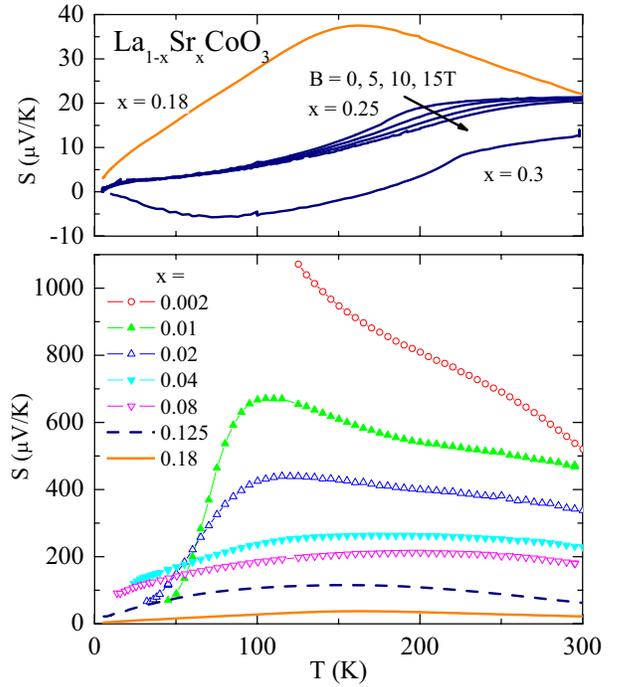}
 \caption[]{Thermopower $S$ of \lsco\ as a function of temperature for
 different doping $x>0$.
 For $x=0.25$ we show $S(T)$ also for different magnetic fields.}
 \label{tk}
\end{figure}

The thermopower measurements of the \lsco\ series for $x>0$ are
presented in Fig.\,\ref{tk}.  For the crystal with the lowest Sr
content $x=0.002$ we find a large positive thermopower which
increases with decreasing temperature. As mentioned above we
could not determine $S$ for highly insulating crystals with $\rho
\gtrsim 10^{9}\,\Omega$cm as it is the case for this crystal below
about 100\,K. With increasing Sr content $S$ systematically
decreases and for $0.01 \le x \le 0.18$ all $S(T)$ curves show
maxima which become less pronounced and slightly shift towards
higher temperature. We note that additional anomalies observed in
the temperature dependence $S(T)$ of polycrystalline \lsco\
(Ref.\,\onlinecite{senaris95b}) are not reproduced by our
single-crystal data. Such a difference has already been observed
in magnetization data, and gives further evidence that these
additional anomalies are not an intrinsic feature of \lsco .

For the metallic samples with $x\ge 0.25$ $S(T)$ varies only
weakly with temperature between 300\,K and $T_c$. Around $T_c$ a
sharp kink occurs and $S$ strongly decreases. For $x=0.3$ there
is even a sign change and $S$ becomes negative for $T<170$\,K.
The sensitivity of $S$ to the magnetic ordering indicates that a
considerable contribution of $S$ arises from magnetic entropy.
Therefore, one may also expect a pronounced magnetic-field
dependence $S(B)$ as has been pointed out recently.\cite{wang03a}
As shown in Fig.\,\ref{tk} we find indeed a strong magnetic-field
induced suppression of $S$, that is most pronounced around $T_c$.
This arises from the fact that the magnetic entropy may be
strongly reduced by available magnetic-field strengths only
around $T_c$. For higher temperatures thermal disorder becomes
too large ($k_BT\gg g\mu_B B$) and for $T\ll T_c$ the magnetic
entropy is already frozen by the magnetic exchange coupling.

At high enough temperatures the thermopower is expected to be
determined by the so-called Heikes formula (see e.\,g.\
Ref.\,\onlinecite{chaikin76a}). This general expression has been
refined for the case of doped cobaltates in
Ref.\,\onlinecite{koshibae00a} to
\begin{equation}
\label{heikes}
S=-\frac{k_B}{e}\left[\ln\left(\frac{n}{1-n}\right) +
\ln\left(\frac{g_3}{g_4}\right)\right] \,\,.
\end{equation}
Here, $n$ denotes the content of Co$^{4+}$ ions and $g_3$ ($g_4$)
is the number of possible configurations of the Co$^{3+}$
(Co$^{4+}$) ions, which is, in general, given by the product of
orbital and spin degeneracy. If different spin states of the
Co$^{3+/4+}$ ions are close enough in energy, the number of
possible configurations may further increase.\cite{koshibae00a}

\begin{figure}
\includegraphics[angle=0,width=8cm]{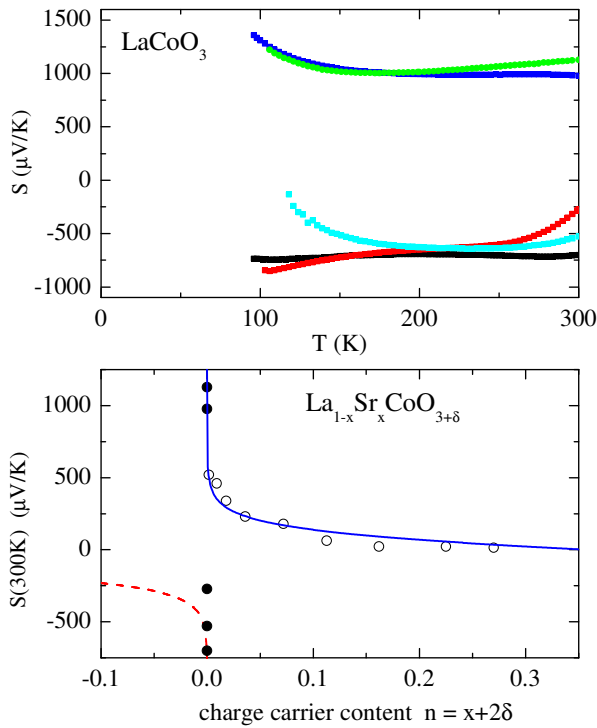}
 \caption[]{Top: Thermopower $S$ as a function of temperature
 for 5 different crystals of nominally undoped \lco . Bottom:
 Room-temperature values $S(300$\,K) as a function of the charge carrier
 concentration $n=x+2\delta$ for Sr-doped ($\circ$) and pure ($\bullet$) \lco .
 The lines are calculated via Eq.\,\ref{heikes}. (see text.)}
 \label{tklco}
\end{figure}

In the lower panel of Fig.\,\ref{tklco} we show the
room-temperature values of $S$ for \lsco\ (open symbols, $x>0$)
as a function of the charge carrier content $n=0.9\cdot x$ as
determined by thermogravimetric analysis (see above). A fit of
the experimental data via Eq.\,\ref{heikes} with the ratio
$g_3/g_4$ as the only fit parameter yields $g_3/g_4\simeq 1.8$.
The saturation magnetization of the samples showing ferromagnetic
order ($x>0.18$) is best described by assuming an $S=1/2$
low-spin state for Co$^{4+}$ and an $S=1$ intermediate-spin state
for Co$^{3+}$.\cite{kriener04a} For this combination of spin
states a ratio $g_3/g_4=3/2$ is expected, which is independent
from the possible orbital degeneracies $\nu=1$ or 3 as long as
both, the Co$^{4+}$ LS and Co$^{3+}$ IS state have the same $\nu
$. If one assumes that the energy of the Co$^{3+}$ LS state is
close to that of the Co$^{3+}$ IS state, the ratio increases to
$g_3/g_4=2$ and 5/3 for $\nu=1$ and 3, respectively. For all of
these cases the experimental data are reasonably well described
by Eq.\,\ref{heikes}. However, this does not exclude other
spin-state combinations, since Eq.\,\ref{heikes} is derived for
the high-temperature limit, whereas in Fig.\,\ref{tklco} the
room-temperature values of $S$ are considered and at least for the
samples with $x<0.25$ the $S(T)$ curves have a more or less
pronounced negative slope at 300\,K. Thus, it is possible that
for higher temperatures a larger $g_3/g_4$ ratio would be
obtained in the fit. In addition, there are also other
combinations of spin states yielding $g_3/g_4$ ratios close to
1.8, but these are not supported by the measured saturation
magnetization. Despite these uncertainties we interpret the
doping dependence of $S(300\,{\rm K})$ as further evidence for
the Co$^{4+}_{LS}$/Co$^{3+}_{IS}$ combination suggested from the
magnetization and resistivity data.\cite{kriener04a}

According to Eq.\,\ref{heikes} the thermopower is expected to
diverge for a vanishing hole content, i.\ e., when the nominally
undoped \lco\ is approached. For electron doping one may still
apply Eq.\,\ref{heikes}, but with a positive sign and $g_4$ for
the degeneracy of Co$^{2+}$. The dashed line in Fig.\,\ref{tklco}
is calculated for $g_4=4$ as expected for Co$^{2+}$ in a HS
state.\cite{blockade} A large negative thermopower has been
observed recently in electron-doped
La$_{1-x}$Ce$_x$CoO$_3$.\cite{maignan04a} For nominally undoped
\lco , we find a different sign of $S$ for different crystals.
This is shown in the upper panel of Fig.\,\ref{tklco}. Although
these crystals have been grown under the same conditions, either a
large negative or a large positive thermopower can be obtained.
We suspect that this extreme sensitivity of $S$ results from weak
deviations in the oxygen content of LaCoO$_{3+\delta}$, which
cause small concentrations $2\delta$ of hole or electron doping
depending on the sign of $\delta$. Already extremely small values
of $|\delta| < 0.002$, which are well below the accuracy of
oxygen determination in cobaltates, would be sufficient to
explain the observed thermopower values with $|S|> 500\,\mu$V/K
in LaCoO$_{3+\delta}$. We also suspect that there are weak
inhomogeneities of the oxygen content in each sample, which can
strongly influence the sign and also the temperature dependence
of the measured thermopower. Most probably, this is also the
reason for the contradictory results for $S(T)$ obtained in
Refs.\,\onlinecite{senaris95a,sehlin95a}. We mention that this
extreme sensitivity $S(\delta)$ is expected only for the (almost)
undoped \lco , whereas for samples with a finite Sr content the
drastic influence of such small variations of $\delta $ rapidly
decreases with increasing $x$.

\subsection{Figure of Merit}

\begin{figure}
\includegraphics[angle=0,width=8cm]{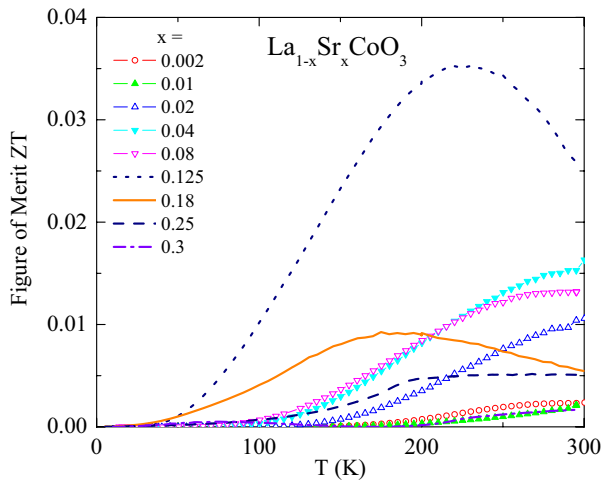}
 \caption[]{Figure of merit of \lsco\ as a function of
 temperature for different
 doping $x$. } \label{merit}
\end{figure}

In Fig.\,\ref{merit} we present the thermoelectric figure of merit
of \lsco . Since the thermal conductivity for all $x$ is rather
low, one precondition for large $ZT$ values is already fulfilled
for the entire \lsco\ series. The very low-doped samples show in
addition very large thermopower values, but their resistivities
are also large and thus prevent large $ZT$ values. Small
resistivities are obtained for large $x$, but these samples also
have smaller thermopower values. Thus the optimum figure of merit
is obtained in the intermediate doping range. We find a maximum
$ZT\simeq 0.035$ around 225\,K, which is as large as the value
observed in \nco .\cite{takahata00a} These values are among the
largest observed in transition metal oxides so far, but are still
too small for technical applications. Nevertheless, it suggests
that cobaltates may be promising candidates for thermoelectric
materials, if further enhancement of $ZT$ is possible. The fact
that Co can occur in different spin states of strongly different
sizes could be used to increase structural and magnetic disorder
in order to suppress the thermal conductivity on the one hand and
enhance the thermopower on the other.

\section{Summary}

We have presented a systematic study of the thermal conductivity
and the thermopower of a series of single crystals of \lscod .
The thermal conductivity is strongly suppressed for the entire
doping range. In pure \lco\ this suppression most probably arises
from local lattice distortions due to a temperature-induced
spin-state transition of the Co$^{3+}$ ions. For small finite
$x$, a spin-state transition of the Co$^{3+}$ ions may be induced
by the neighboring magnetic Co$^{4+}$ ions, and so-called
high-spin polarons can be formed, which also cause lattice
disorder. For larger $x$ this effect becomes less important, but
scattering of phonons by mobile charge carriers will also play a
role for the suppression of the thermal conductivity. For finite
doping we find a large, positive thermopower, which strongly
depends on temperature and doping. The room temperature values of
the thermopower follow a doping dependence that is expected from a
modified Heikes formula,\,\cite{koshibae00a} if we  assume an
intermediate-spin state for Co$^{3+}$ and a low-spin state for
Co$^{4+}$ as it is suggested from magnetization and resistivity
data.\cite{kriener04a} In nominally undoped \lco\ we find for
different crystals either a large positive or a large negative
thermopower. We suspect that this is a consequence of weak
deviations ($|\delta | < 0.002$) from the nominal oxygen content
causing small amounts of hole or electron doping. Most probably,
a weak oxygen off-stoichiometry is also the reason for the
contradictory results of the thermopower reported for \lco\
previously.\cite{senaris95a,sehlin95a} For the crystal with
$x=0.25$ both, the thermal conductivity and the thermopower show
a significant magnetic-field dependence in the temperature range
around the ferromagnetic ordering temperature. The field
dependence of the thermal conductivity can be traced back to a
field-dependent charge carrier contribution to the heat current,
and the field dependence of the thermopower indicates that it
contains a sizeable contribution arising from magnetic entropy.
From the resistivity $\rho$, the thermopower $S$ and the thermal
conductivity $\kappa$ we have also determined the thermoelectric
figure of merit $Z=S^2\,/\,\kappa\rho$, which strongly depends on
both, doping and temperature.  A maximum of $ZT \simeq 0.035$ is
obtained  for $x=0.125$ and $200\, {\rm K}\le T\le 250$\,K. This
value is large, but yet too small for technical applications.
Nevertheless, it indicates that Co-based materials could be
interesting candidates for thermoelectric cooling.

\begin{acknowledgments}
We thank S.\ Heijligen for technical assistance during the
measurements of the thermogravimetric analysis. We acknowledge
financial support by the Deutsche Forschungsgemeinschaft through
SFB\,608.
\end{acknowledgments}


\begin{thebibliography}{10}
\parskip-0.2ex plus0.05ex minus0.05ex

\bibitem{jonker53a}
G.~H. Jonker and J.~H.~Van Santen.
\newblock Physica {\bf XIX}, 120 (1953).

\bibitem{asai94a}
K.~Asai, O.~Yokokura, N.~Nishimori, H.~Chou, J.M. Tranquada,
G.~Shirane,
  S.~Higuchi, Y.~Okajima, and K.~Kohn.
\newblock Phys.\ Rev.\ B {\bf 50}, 3025 (1994).

\bibitem{senaris95a}
M.~A.~Se\~{n}ar\'{\i}s Rodr\'{\i}guez and J.~B. Goodenough.
\newblock J.\ of Solid State Chem. {\bf 116}, 224 (1995).

\bibitem{korotin96a}
M.~A. Korotin, S.~Y. Ezhov, I.~V. Solovyev, V.~I. Anisimov, D.~I.
Khomskii, and
  G.~A. Sawatzky.
\newblock Phys.\ Rev.\ B {\bf 54}, 5309 (1996).

\bibitem{yamaguchi97a}
S.~Yamaguchi, Y.~Okimoto, and Y.~Tokura.
\newblock Phys.\ Rev.\ B {\bf 55}, 8666 (1997).

\bibitem{asai98a}
K.~Asai, A.~Yoneda, O.~Yokokura, J.M. Tranquada, G.~Shirane, and
K.~Kohn.
\newblock J.\ Phys.\ Soc.\ Japan {\bf 67}, 290 (1998).

\bibitem{zobel02a}
C.~Zobel, M.~Kriener, D.~Bruns, J.~Baier, M.~Gr\"uninger,
T.~Lorenz,
  P.~Reutler, and A.~Revcolevschi.
\newblock Phys.\ Rev.\ B {\bf 66}, 020402 (2002).

\bibitem{vogt00a}
T.~Vogt, P.M. Woodward, P.~Karen, B.A. Hunter, P.~Henning, and
A.R.
  Moodenbaugh.
\newblock Phys.\ Rev.\ Lett. {\bf 84}, 2969 (2000).

\bibitem{suard00a}
E.~Suard, F.~Fauth, V.~Caignaert, I.~Mirebeau, and G.~Baldinozzi.
\newblock Phys.\ Rev.\ B {\bf 61}, 11871 (2000).

\bibitem{moritomo00a}
Y.~Moritomo, T.~Akimoto, M.~Takeo, A.~Machida, E.~Nishibori,
M.~Takata,
  M.~Sakata, K.~Ohoyama, and A.~Nakamura.
\newblock Phys.\ Rev.\ B {\bf 61}, R13325 (2000).

\bibitem{respaud01a}
M.~Respaud, C.~Frontera, J.L. Garc\'{\i}a-Mu{\~{n}}oz, M.A.G.
Aranda,
  B.~Raquet, J.M. Broto, H.~Rakoto, M.~Goiran, A.~Llobet, and
  J.~Rodr\'{\i}guez-Carvajal.
\newblock Phys.\ Rev.\ B {\bf 64}, 214401 (2001).

\bibitem{wu01a}
H.~Wu.
\newblock Phys.\ Rev.\ B {\bf 64}, 92413 (2001).

\bibitem{wang01a}
J.~Wang, Weiyi Zhang, and D.~Y. Xing.
\newblock Phys.\ Rev.\ B {\bf 64}, 64418 (2001).

\bibitem{roy02a}
S.~Roy, M.~Khan, Y.Q. Guo, J.~Craig, and N.~Ali.
\newblock Phys.\ Rev.\ B {\bf 65}, 64437 (2002).

\bibitem{hu04a}
Z.~Hu, H.~Wu, M.~W. Haverkort, H.~H. Hsieh, H.-J. Lin, T.~Lorenz,
J.~Baier,
  A.~Reichl, I.~Bonn, C.~Felser, A.~Tanaka, C.~T. Chen, and L.~H. Tjeng.
\newblock Phys.\ Rev.\ Lett. {\bf 92}, 207402 (2004).

\bibitem{takada03a}
K.~Takada, H.~Sakurai, E.~Takayama-Muromachi, F.~Izumi, R.A.
Dilanian, and
  T.~Sasaki.
\newblock Nature {\bf 422}, 53 (2003).

\bibitem{takahata00a}
K.~Takahata, Y.~Iguchi, D.~Tanaka, T.~Itoh, and I.~Terasaki.
\newblock Phys.\ Rev.\ B {\bf 61}, 12551 (2000).

\bibitem{venkata_nature01a}
R.~Venkatasubramanian, E.~Siivola, T.~Colpitts, and B.~O'Quinn.
\newblock Nature {\bf 413}, 597 (2001).

\bibitem{harmann_science02a}
T.C. Harman, P.J. Taylor, M.P. Walsh, and B.E. LaForge.
\newblock Science {\bf 297}, 2229 (2002).

\bibitem{terasaki97a}
I.~Terasaki, Y.~Sasago, and K.~Uchinokura.
\newblock Phys.\ Rev.\ B {\bf 56}, R12685 (1997).

\bibitem{wang03a}
Y.~Wang, N.S. Rogado, R.J. Cava, and N.P. Ong.
\newblock Nature {\bf 423}, 425 (2003).

\bibitem{senaris95b}
M.A.~Se\~{n}ar\'{\i}s Rodr\'{\i}guez and J.B. Goodenough.
\newblock J.\ of Solid State Chem. {\bf 118}, 323 (1995).

\bibitem{anilkumar98a}
P.S.~Anil Kumar, P.A. Joy, and S.K. Date.
\newblock J.\ Appl.\ Phys. {\bf 83}, 7375 (1998).

\bibitem{itoh94a}
M.~Itoh, I.~Natori, S.~Kubota, and K.~Motoya.
\newblock J.\ Phys.\ Soc.\ Japan {\bf 63}, 1486 (1994).

\bibitem{kriener04a}
M.~Kriener, C.~Zobel, A.~Reichl, J.~Baier, M.~Cwik, K.~Berggold,
H.~Kierspel,
  O.~Zabara, A.~Freimuth, and T.~Lorenz.
\newblock Phys.\ Rev.\ B {\bf 69}, 094417 (2004).

\bibitem{sehlin95a}
S.R. Sehlin, H.U. Anderson, and D.M. Sparlin.
\newblock Phys.\ Rev.\ B {\bf 52}, 11681 (1995).

\bibitem{baier05a}
J.~Baier, S.~Jodlauk, M.~Kriener, A.~Reichl, C.~Zobel,
H.~Kierspel,
  A.~Freimuth, and T.~Lorenz.
\newblock Phys.\ Rev.\ B {\bf 71}, 014443 (2005).

\bibitem{lengsdorf04a}
R.~Lengsdorf, M.~Ait-Tahar, S.S. Saxena, M.~Ellerby, D.I.
Khomskii,
  H.~Micklitz, T.~Lorenz, and M.M. Abd-Elmeguid.
\newblock Phys.\ Rev.\ B {\bf 69}, 140403(R) (2004).

\bibitem{remarkO2p}
In the case of hole doping this formal description must not be
taken literally,
  because due to the high oxidation state Co$^{4+}$ the holes are likely to
  enter the oxygen $2p$ states.

\bibitem{blockade}
Note that depending on the spin states of the various Co ions
hole and electron
  doping may be strongly different with respect to charge transport due to the
  so-called spin blockade effect\,\cite{maignan04a,maignan04b}.

\bibitem{james04a}
M.~James, D.~Cassidy, D.J. Goossens, and R.L. Withers.
\newblock J.\ of Solid State Chem. {\bf 177}, 1886--1895 (2004).

\bibitem{goossens04a}
D.J. Goossens, K.F. Wilson, M.~James, A.J. Studer, and X.L. Wang.
\newblock Phys.\ Rev.\ B {\bf 69}, 134411 (2004).

\bibitem{conder05a}
K.~Conder, E.~Pomjakushina, A.~Soldatov, and E.~Mitberg.
\newblock Mater.\ Res.\ Bull. {\bf 40}, 257--263 (2005).

\bibitem{verweis_scbo}
Resonant scattering may also arise from scattering by magnetic
excitations as
  it is observed in SrCu$_2$(BO$_3$)$_2$\,(Ref.\,\onlinecite{hofmann02a}).

\bibitem{yan04a}
J.-Q. Yan, J.-S. Zhou, and J.B. Goodenough.
\newblock Phys.\ Rev.\ B {\bf 69}, 134409 (2004).

\bibitem{yamaguchi96a}
S.~Yamaguchi, Y.~Okimoto, H.~Taniguchi, and Y.~Tokura.
\newblock Phys.\ Rev.\ B {\bf 53}, 2926 (1996).

\bibitem{mahendiran96a}
R.~Mahendiran and A.K. Raychaudhuri.
\newblock Phys.\ Rev.\ B {\bf 54}, 16044 (1996).

\bibitem{chaikin76a}
P.M. Chaikin and G.~Beni.
\newblock Phys.\ Rev.\ B {\bf 13}, 647 (1976).

\bibitem{koshibae00a}
W.~Koshibae, K.~Tsutsui, and S.~Maekawa.
\newblock Phys.\ Rev.\ B {\bf 62}, 6869 (2000).

\bibitem{maignan04a}
A.~Maignan, D.~Flahaut, and S.~H\'{e}bert.
\newblock The European Phys.\ J.\ B {\bf 39}, 145--148 (2004).

\bibitem{maignan04b}
A.~Maignan, V.~Caignaert, B.~Raveau, D.~Khomskii, and G.~Sawatzky.
\newblock Phys.\ Rev.\ Lett. {\bf 93}, 026401 (2004).

\bibitem{hofmann02a}
M.~Hofmann, T.~Lorenz, G.S. Uhrig, H.~Kierspel, O.~Zabara,
A.~Freimuth,
  H.~Kageyama, and Y.~Ueda.
\newblock Phys.\ Rev.\ Lett. {\bf 87}, 047202 (2001).

\end{thebibliography}

\end{document}